\documentclass[conference,a4paper]{IEEEtran}

\IEEEoverridecommandlockouts

\ifCLASSINFOpdf
   \usepackage[pdftex]{graphicx}
\else
   \usepackage[dvips]{graphicx}
\fi
\usepackage[cmex10]{amsmath}
\usepackage{amssymb}
\usepackage{algorithmic}

\usepackage{algorithm}

\newtheorem{theorem}{Theorem}
\newtheorem{corollary}{Corollary}

\begin{document}
%
\title{Link Scheduling in STDMA Wireless Networks: A Line Graph Approach}

\author{\IEEEauthorblockN{N. Praneeth Kumar}
\IEEEauthorblockA{Goldman Sachs (India) Securities Pvt. Ltd.\\
EGL Business Park, Domlur, Bangalore, India\\
Email: praneeth.kumar@gs.com}
\and
\IEEEauthorblockN{Ashutosh Deepak Gore and Abhay Karandikar}
\IEEEauthorblockA{Department of Electrical Engineering\\
Indian Institute of Technology - Bombay, India\\
Email: \{adgore,karandi\}@ee.iitb.ac.in}
}


%


\maketitle

\begin{abstract}
  We consider point to point link scheduling in Spatial Time Division
  Multiple Access (STDMA) wireless networks under the physical
  interference model.  We propose a novel link scheduling algorithm
  based on a line graph representation of the network, by embedding
  the interferences between pairs of nodes into the edge weights of
  the line graph. Our algorithm achieves lower schedule length and
  lower run time complexity than existing algorithms.
\end{abstract}


%
\IEEEpeerreviewmaketitle

\section{Introduction}

A prevalent scheme for spatial reuse in wireless networks is Spatial
Time Division Multiple Access (STDMA)
\cite{nelson_kleinrock__spatial_tdma}, in which time is divided into
fixed-length slots that are organized cyclically. An STDMA link
schedule defines the transmission rights for each time slot in such a
way that communicating pairs of nodes, i.e., links, assigned to the
same time slot do not collide.  The interference between the links can
be modeled by assuming that transmission on a link is successful if
the distance between the nodes is less than some {\em communication
  range} and no other node is transmitting within an {\em interference
  range} from the receiver of the link. This is called protocol
interference model \cite{gupta_kumar__capacity_wireless}. Under these
assumptions, the network can be modeled by a {\em communication graph}
and scheduling algorithms employ edge coloring techniques to minimize
the schedule length. Though determining an optimal schedule is known
to be NP-complete \cite{ramanathan_lloyd__scheduling_algorithms},
heuristics have been proposed in
\cite{ramanathan_lloyd__scheduling_algorithms}, 
\cite{wang_li_song__efficient_interference}, 
\cite{alicherry_bhatia__joint_channel}.

A more realistic model would be to consider a transmission on a link
to be successful if the Signal to Interference and Noise Ratio (SINR)
at the receiver is greater than some threshold, say $\gamma_c$. This
model is called physical interference model
\cite{gupta_kumar__capacity_wireless}.  Recently, few authors
\cite{brar_blough_santi__computationally_efficient}, 
\cite{moscibroda_wattenhofer__complexity_connectivity}, 
\cite{gore_karandikar__link_scheduling}, 
\cite{gore_karandikar__high_spatial} have proposed scheduling
algorithms based on this physical interference model, which result in
improved network throughput. In
\cite{brar_blough_santi__computationally_efficient}, the authors
propose a polynomial time algorithm which gives a provable performance
guarantee within a constant factor compared to the optimal.  This
algorithm is a greedy heuristic that determines a schedule based on
the feasibility of satisfying SINR conditions using a communication
graph based representation of the network.

In this paper, we propose a novel scheduling algorithm for STDMA
wireless networks under physical interference model.  Our approach is
based on a line graph \cite{west__graph_theory} representation of the
network where the weights of the edges correspond to interferences
between pairs of nodes.  Analogous to a line graph, a conflict graph
model under physical interference assumptions has been suggested in
\cite{jain_padhye__impact_interference}.  However, the authors of
\cite{jain_padhye__impact_interference} do not propose any specific
scheduling algorithm and use the weighted conflict graph only to
compute bounds on the network throughput.  On the other hand, we use a
line graph representation of the network under the physical
interference model and develop a novel scheduling algorithm with lower
time complexity and substantially improved performance in terms of
schedule length.

The rest of the paper is organized as follows. In Section \ref{tejkd},
we explain our system model, formulate the problem and describe the
proposed algorithm. In Section \ref{yuydk}, we prove the correctness
of our algorithm and derive its computational complexity.  The
performance of our algorithm is evaluated in Section \ref{simul_sec}.
We conclude in Section \ref{conc}.

\section{Line graph based link scheduling algorithm}
\label{tejkd}

Consider an STDMA wireless network $\Upsilon(\cdot)$ with $N$ static
nodes (wireless routers) in a two-dimensional plane.  During a time
slot, a node can transmit to exactly one node, receive from exactly
one node or remain idle.  We assume homogeneous and backlogged
nodes. Let:
\begin{eqnarray*}
P &=& \mbox{transmission power of every node}\\
N_0 &=& \mbox{thermal noise power spectral density}\\
\alpha &=& \mbox{path loss factor}
\end{eqnarray*}
The received signal power at a distance $d$ from a transmitter is
given by $\frac{P}{d^{\alpha}}$. The STDMA wireless network
$\Upsilon(\cdot)$ is modeled by its communication graph $\mathcal
G(\mathcal V, \mathcal E)$, where $\mathcal V$ and $\mathcal E$ denote
the sets of vertices and edges respectively, as follows:
\begin{enumerate}
\item 
  Every node $i$ in $\Upsilon(\cdot)$ is represented by a vertex $v_i
  \in \mathcal V$ in $\mathcal G(\cdot)$.

\item
  If the Euclidean distance between two distinct nodes $a$ and $b$ is
  no greater than the communication range $R_c := (\frac{P}{N_0
    \gamma_c})^{\frac{1}{\alpha}}$, then there is a directed edge from
  $v_a$ to $v_b$ in $\mathcal G(\cdot)$, i.e., $(v_a,v_b) \in
  \mathcal E$.
\end{enumerate}
The set of edges in $\mathcal G(\cdot)$ to be scheduled is determined
by a routing algorithm.  For simplicity, we only consider exhaustive
schedules, i.e., schedules which assign exactly one time slot to every
directed edge in $\mathcal G(\cdot)$.

\begin{algorithm}[h!]
\caption{LineGraphLinkSchedule (LGLS)}
\label{line_graph}
\begin{algorithmic}[1]
\STATE{\bf Input:} Communication graph ${\mathcal G}$(${\mathcal V}$,${\mathcal E}$), $\gamma_c$, $N_0$, $P$
\STATE{\bf Output:} A coloring ${\mathcal C}$: ${\mathcal E} \rightarrow \{1,2,\ldots\}$ 
\STATE ${\mathcal V'} \leftarrow {\mathcal E}$ \COMMENT{Phase 1 begins}
\STATE $\mbox{Construct the directed complete graph }{\mathcal G'}({\mathcal V'},{\mathcal E'})$
\FORALL{$e'_{ij} \in {\mathcal E'}$}
\IF{$\mbox{edges } i \mbox{ and } j \mbox{ have a common vertex in } {\mathcal G(\cdot)}$}
\STATE $w_{ij} \leftarrow 1$
\ELSE
\STATE $w_{ij} \leftarrow \gamma_c \frac{{d(t_j,r_j)}^\alpha}{{d(t_i,r_j)}^\alpha}$
\ENDIF
\ENDFOR \COMMENT{Phase 1 ends}
\FORALL[Phase 2 begins]{$e'_{ij} \in {\mathcal E'}$}
\STATE $w'(e'_{ij}) \leftarrow \max\{0,1-w_{ij}\}$
\ENDFOR \COMMENT{Phase 2 ends}
\FORALL[Phase 3 begins]{$v_j' \in {\mathcal V'}$}
\STATE ${\mathcal N}(v_j') \leftarrow \frac{N_0\gamma_c}{P} {{d(t_j,r_j)}^\alpha}$
\ENDFOR \COMMENT{Phase 3 ends}
\STATE ${\mathcal V_{uc}'} \leftarrow {\mathcal V'}$ \COMMENT{Phase 4 begins}
\STATE $p \leftarrow 0$
\WHILE {${\mathcal V_{uc}'} \neq \phi$}
\STATE $p \leftarrow p+1$
\STATE $\mbox{choose } v' \in {\mathcal V_{uc}'}$ randomly
\STATE ${\mathcal C}(v') \leftarrow p$
\STATE ${\mathcal V_{uc}'} \leftarrow {\mathcal V_{uc}'} \setminus \{v'\}$
\STATE ${\mathcal V_p'} \leftarrow \{v'\}$
\STATE $\psi \leftarrow \mbox{1}$
\WHILE {$\psi=1 \mbox{ and } {\mathcal V_{uc}'} \neq \phi$}
\STATE $u' \leftarrow \arg\max_{y' \in {\mathcal V}_{uc}'} \sum_{x' \in {\mathcal V}_p'} w'(e'_{x'y'})+w'(e'_{y'x'})$
\FORALL {$v_c' \in {\mathcal V_p'}$}
\IF {$\sum_{v_1' \in {\mathcal V_p'} \setminus \{v_c'\} \cup \{u'\}} w'(e'_{{v_1'}{v_c'}}) \leqslant |{\mathcal V_p'}|+{\mathcal N}(v_c')-1$}
\STATE $\psi \leftarrow 0$
\ENDIF
\ENDFOR
\IF {$\psi=1 \mbox{ and } \sum_{v_1' \in {\mathcal V_p'}} w'(e'_{{v_1'}u'}) > |{\mathcal V_p'}|+{\mathcal N}(u')-1$}
\STATE ${\mathcal C}(u') \leftarrow p$
\STATE ${\mathcal V_p'} \leftarrow {\mathcal V_p'} \cup \{u'\}$
\STATE ${\mathcal V_{uc}'} \leftarrow {\mathcal V_{uc}'} \setminus \{u'\}$
\ELSE
\STATE $\psi \leftarrow 0$
\ENDIF
\ENDWHILE
\ENDWHILE \COMMENT{Phase 4 ends}
\end{algorithmic}
\end{algorithm}
We now motivate our Line Graph based Link Scheduling (LGLS) algorithm,
and provide the pseudocode in Algorithm \ref{line_graph}.  In Phase
1, we first construct a directed complete line graph
\cite{west__graph_theory} ${\mathcal G'}({\mathcal V'},{\mathcal E'})$
which has the edges of ${\mathcal G}(\cdot)$ as its vertices, i.e.,
${\mathcal V'}={\mathcal E}$.  Let the edges of ${\mathcal G}(\cdot)$
and the corresponding vertices of ${\mathcal G'}(\cdot)$ be labeled
$1,2,\ldots,e$. Let $t_i$ and $r_i$ denote the transmitter and
receiver respectively of edge $i$ in ${\mathcal G}(\cdot)$. Let
$d(t_i,r_j)$ denote the Euclidean distance between $t_i$ and $r_j$.

For any two edges $i$ and $j$ in graph ${\mathcal G}(\cdot)$, the
weight function $w_{ij}$ is defined as:
\begin{equation}
\begin{split}
\nonumber
w_{ij}&= 
\begin{cases} 
1 & \text{if } i \text{ and } j 
 \text{ have a common vertex}, \\
\gamma_c \frac{{d(t_j,r_j)}^\alpha}{{d(t_i,r_j)}^\alpha} & \text{otherwise.}
\end{cases}
\end{split}
\end{equation}
This weight function $w_{ij}$ indicates the interference energy at
$r_j$ due to transmission from $t_i$ to $r_i$ scaled with respect to
the signal energy of $t_j$ at $r_j$.

In Phase 2, we compute the co-schedulability weight function
$w'(\cdot)$. For any two edges $i$ and $j$ in $\mathcal G(\cdot)$, the
weight of the edge $e'_{ij}$ in ${\mathcal G'}(\cdot)$ is given by
$w'(e'_{ij})=\max\{0,1-w_{ij}\}$.  Since $w_{ij}$ and $w_{ji}$ represent
the interference between links $i$ and $j$ in $\Upsilon(\cdot)$,
$w'(e'_{ij})$ and $w'(e'_{ji})$ intuitively represent the
co-schedulability of vertices $i$ and $j$ in $\mathcal G'(\cdot)$.
For example, if $w_{ij}$ is greater than or equal to $1$, it means
that the interference between links $i$ and $j$ in $\Upsilon(\cdot)$
is very high and these links cannot be scheduled simultaneously. This
will result in $w'(e'_{ij})$ being equal to $0$ indicating that the
vertices $i$ and $j$ in $\mathcal G'(\cdot)$ 
are not co-schedulable.  On the other hand, if
$w_{ij}$ is almost $0$, $w'(e'_{ij})$ will be almost $1$ indicating
that the vertices $i$ and $j$ in $\mathcal G'(\cdot)$ are
co-schedulable. 
In Phase 3, we determine the normalized noise power at the receiver of
each vertex of ${\mathcal G'}(\cdot)$.

Our objective is to color the vertices of ${\mathcal
G'}(\cdot)$ (equivalently, edges of $\mathcal G(\cdot)$) using minimum
number of colors under the physical interference model, i.e., subject
to the condition that the SINR at every receiver of each link in
$\Upsilon(\cdot)$ is greater than the communication threshold
$\gamma_c$.  Equivalently, for any ${\mathcal V_{cc}'} \subseteq
{\mathcal V'}$, the coloring of all nodes $v_i' \in {\mathcal
V_{cc}'}$ with the same color is defined to be {\em feasible} if
\begin{eqnarray}
\frac{\frac{P}{{d(t_{v_i'},r_{v_i'})}^\alpha}}{N_0+\sum_{v_j'\in{\mathcal
V_{cc}'} \setminus \{v_i'\}}\frac{P}{{d(t_{v_j'},r_{v_i'})}^\alpha}}
\geqslant \gamma_c \;\forall\; v_i' \in {\mathcal V_{cc}'}
\label{qwerty}
\end{eqnarray}
In $\mathcal G'(\cdot)$, this condition translates to the sum of
weights of edges incoming to a vertex from all the co-colored vertices
being greater than the sum of the number of {\em remaining} co-colored
vertices and the normalized noise power minus a constant factor (unity).

The actual coloring of vertices of $\mathcal G'(\cdot)$, i.e., edges
of $\mathcal G(\cdot)$, occurs in Phase 4.  Let ${\mathcal V_{uc}'}$
at any time denote the set of uncolored vertices of $\mathcal
G'(\cdot)$ till that time.  Initially, $\mathcal V_{uc}'$ includes all
the vertices of $\mathcal G'(\cdot)$.  First we choose a vertex
randomly from ${\mathcal V_{uc}'}$. This is assigned a new color, let
it be $p$. Then we choose that vertex from ${\mathcal V_{uc}'}$ which
maximizes the sum of weights of all the edges between that vertex and
the vertices colored by $p$.  Now for each vertex colored with $p$, we
check if the sum of weights of all the incoming edges is greater than
the sum of the number of vertices colored with $p$ and the normalized
noise power minus a constant factor (unity).  If it is satisfied, the
vertex is colored with $p$. If not, it is colored with a new
color. The algorithm exits when all the vertices are colored.

\section{Analysis of LGLS Algorithm}
\label{yuydk}

In this section, we prove the correctness of the LGLS algorithm and
derive its run time (computational) complexity. We follow the notation
of Algorithm \ref{line_graph}.

\begin{theorem}
  For any ${\mathcal V_{cc}'} \subseteq {\mathcal V'}$, if
  $\sum_{v_2'\in{\mathcal V_{cc}'} \setminus \{v_1'\}}
  w'(e'_{v_2'v_1'})$ $> |{\mathcal V_{cc}'}|+{\mathcal N}(v_1')-2$
  $\;\forall\; v_1' \in {\mathcal V_{cc}'}$, then the coloring of all the
  vertices of ${\mathcal V_{cc}'}$ with the same color is feasible.
\label{correctness}
\end{theorem}
\begin{IEEEproof}
  $w'(e'_{v_2'v_1'}) \in [0,1]$.  Suppose $w'(e'_{v_3'v_1'})=0$ for
  some $v_1', v_3' \in {\mathcal V_{cc}'}$, $v_1' \neq v_3'$, then
  $\sum_{v_2'\in{\mathcal V_{cc}'} \setminus \{v_1'\}}
  w'(e'_{v_2'v_1'})=$ $\sum_{v_2'\in{\mathcal V_{cc}'} \setminus
    \{v_1',v_3'\} } w'(e'_{v_2'v_1'})\leqslant$ $\sum_{v_2'\in{\mathcal
      V_{cc}'} \setminus \{v_1',v_3'\} } 1 =$ $|{\mathcal V_{cc}'}
  \setminus \{v_1',v_3'\}|=|{\mathcal V_{cc}'}|-2$, which contradicts
  the hypothesis. So, an edge connecting any two vertices in
  ${\mathcal V_{cc}'}$ has positive weight. But, $w'(e'_{v_2'v_1'})=0
  $ or $ 1-w_{v_2'v_1'}$.  So, $w'(e'_{v_2'v_1'})=1-w_{v_2'v_1'}
  \;\forall\; v_1', v_2' \in {\mathcal V_{cc}'}$, $v_1' \neq v_2'$ and
  $w_{v_2'v_1'}<1 \;\forall\; v_1', v_2' \in {\mathcal V_{cc}'}$,
  $v_1' \neq v_2'$.  If two vertices $v_1', v_2' \in {\mathcal
    V_{cc}'}$ have a common vertex in $\mathcal G(\cdot)$, then
  $w_{v_2'v_1'}=1$, which is a contradiction. So no two vertices in
  ${\mathcal V_{cc}'}$ have a
  common vertex in $\mathcal G(\cdot)$.  From the hypothesis, \\
  $\sum_{v_2'\in{\mathcal V_{cc}'} \setminus \{v_1'\}}
  w'(e'_{v_2'v_1'}) > |{\mathcal V_{cc}'}|+{\mathcal N}(v_1')-2$ $
  \;\forall\; v_1' \in {\mathcal V_{cc}'}$ \\ $\Leftrightarrow
  \sum_{v_2'\in{\mathcal V_{cc}'} \setminus \{v_1'\}} (1-w_{v_2'v_1'})
  > |{\mathcal V_{cc}'}|+{\mathcal N}(v_1')-2$ \\ $\Leftrightarrow
  |{\mathcal V_{cc}'} \setminus \{v_1'\}|-\sum_{v_2'\in{\mathcal
      V_{cc}'} \setminus \{v_1'\}}w_{v_2'v_1'} > $ $|{\mathcal
    V_{cc}'}|+{\mathcal N}(v_1')-2$ \\ $\Leftrightarrow |{\mathcal
    V_{cc}'}|-1-\sum_{v_2'\in{\mathcal V_{cc}'} \setminus
    \{v_1'\}}w_{v_2'v_1'} > $ $|{\mathcal V_{cc}'}|+{\mathcal
    N}(v_1')-2$ \\ $\Leftrightarrow \sum_{v_2'\in{\mathcal V_{cc}'}
    \setminus \{v_1'\}}w_{v_2'v_1'} + {\mathcal N}(v_1') < 1 $ \\
  $\Leftrightarrow \sum_{v_2'\in{\mathcal V_{cc}'} \setminus \{v_1'\}}
  \gamma_c
  \frac{{d(t_{v_1'},r_{v_1'})}^\alpha}{{d(t_{v_2'},r_{v_1'})}^\alpha}
  + $ $\frac{N_0 \gamma_c}{P}{d(t_{v_1'},r_{v_1'})}^\alpha < 1$ \\
  $\Leftrightarrow
  \frac{\frac{P}{{d(t_{v_1'},r_{v_1'})}^\alpha}}{N_0+\sum_{v_2'\in{\mathcal
        V_{cc}'} \setminus
      \{v_1'\}}\frac{P}{{d(t_{v_2'},r_{v_1'})}^\alpha}}>\gamma_c$
  $\;\forall\; v_1' \in {\mathcal V_{cc}'}$. \\
  Therefore, the SINR threshold condition (\ref{qwerty}) is satisfied
  at all the receivers of all the vertices of ${\mathcal V_{cc}'}$.
\end{IEEEproof}
\vspace{1ex}
With respect to (w.r.t.) the communication graph $\mathcal G(\mathcal
V,\mathcal E)$, let:
\begin{eqnarray*}
e &=& \mbox{number of  edges}\\
v &=& \mbox{number of vertices}
\end{eqnarray*}

\begin{theorem}
  The run time complexity of LGLS algorithm under uniform load
  conditions is $O(e^2)$.
\label{unfrm}
\end{theorem}
\begin{IEEEproof}
  $|{\mathcal V'}|=|{\mathcal E}|$. Since $\mathcal G'(\cdot)$ is a
  directed complete graph, $|{\mathcal E'}|=|{\mathcal V'}|^2=e^2$.
  Since the computation of the function $w_{ij}$ for a given $i$ and
  $j$ takes unit time, the computation of $w_{ij}$ for all edges $i$
  and $j$ of $\mathcal G(\cdot)$ takes $O(e^2)$ time. Hence, the time
  complexity of Phase 1 is $O(e^2)$.  Similarly, the time complexity
  of Phase 2 is also $O(e^2)$.  The time complexity of Phase 3 is
  $O(e)$.  In $\mathcal G'(\cdot)$, let $C$ denote the total number of
  colors used to color all vertices, and let $N_i$ denote the number
  of vertices assigned color $i$.  Since $C$ can never exceed the
  number of vertices in $\mathcal G'(\cdot)$ (the number of edges in
  $\mathcal G(\cdot)$), $C$ is $O(e)$.  The time required by Lines
  21-26 is $O(1)$, let it be $k_1$, where $k_1$ is a constant.  By
  using a careful implementation of storing $\sum_{x' \in {\mathcal
      V}_p'} w'(e'_{x'y'})$ and $\sum_{x' \in {\mathcal V}_p'}
  w'(e'_{y'x'}) \;\forall\; y' \in {\mathcal V_{uc}'}$, the time
  required by Line 28 will be $O(|{\mathcal V_{uc}'}|)$. Let it be
  equal to $a_1 |{\mathcal V_{uc}'}|$, where $a_1$ is a constant.
  With a careful implementation of storing $\sum_{x' \in {\mathcal
      V}_p' \setminus \{y'\} \cup \{u'\}} w'(e'_{x'y'}) \;\forall\; y'
  \in {\mathcal V_p'}$, Lines 30-32 take $O(1)$ time.  Hence, Lines
  29-33 take time $O(|{\mathcal V}_p'|)$, let it be $a_2 |{\mathcal
    V_p'}|$, where $a_2$ is a constant.  Lines 34-40 take
  $O(|{\mathcal V}_p'|)$ time, let it be $a_3 |{\mathcal V_p'}|$,
  where $a_3$ is a constant.  Thus the total run time of Phase 4 is,\\
  $\tau = \sum_{i=1}^C (k_1 + \sum_{m=1}^{N_i} (a_1 |{\mathcal
    V_{uc}'}|+(a_2+a_3) |{\mathcal V_p'}|))$.
   \begin{eqnarray*}
{\mathcal V_{uc}'},{\mathcal V_p'} \subseteq {\mathcal V'}
  &\Rightarrow& 
     |{\mathcal V_{uc}'}|,|{\mathcal V_p'}| \leqslant |{\mathcal V'}|=e.\\
   \Rightarrow \tau 
     &\leqslant& \sum_{i=1}^C \Big(k_1 + \sum_{m=1}^{N_i} (a_1+a_2+a_3)e\Big) \\
     &=& \sum_{i=1}^C k_1 + (a_1+a_2+a_3)e \sum_{i=1}^C \sum_{m=1}^{N_i} 1 \\
     &=& k_1C+(a_1+a_2+a_3)e(e) \\
     &=& O(e^2)
   \end{eqnarray*}
  Hence, the total time complexity of Algorithm \ref{line_graph} is
  $O(e^2)$.
\end{IEEEproof}

We compare our work with that of
\cite{brar_blough_santi__computationally_efficient}, since it is the
only work in STDMA link scheduling whose system model is closest to
our system model.

Under uniform load on all edges, the time complexity of GreedyPhysical
(GP) algorithm of \cite{brar_blough_santi__computationally_efficient}
is $O(ve^{2})$, whereas the time complexity of the proposed LGLS
algorithm is $O(e^2)$. Thus, under uniform load conditions, the time
complexity of LGLS is much lower than that of GP.

Note that our algorithm can be easily extended to incorporate
non-uniform load conditions by assigning integer weights to edges in
$\mathcal G(\cdot)$ and then considering each edge split into those
many edges.  Similar to
\cite{brar_blough_santi__computationally_efficient}, we assume that
the ratio of the maximum load at any node to the minimum load at any
node is upper bounded by a constant.
\begin{corollary}
  The run time complexity of LGLS algorithm under non-uniform load
  conditions is $O(v^2e^2)$.
\end{corollary}
\begin{IEEEproof}
  Let $l_{max}$ and $l_{min}$ denote the maximum load and minimum load
  offered by a node in $\mathcal G(\cdot)$. We assume that
  $\frac{l_{max}}{l_{min}}$ is upper bounded by $k$, a constant.  We
  normalize the load at each node w.r.t. $l_{min}$.  So, the load at
  each node is $i$, where $1 \leqslant i \leqslant k$. Thus, the
  weight on each edge in $\mathcal G(\cdot)$ is upper bounded by
  $\sum_{v \in \mathcal V} i \leqslant vk$, which is $O(v)$. Hence,
  the total number of edges after replacing an edge by the number of
  edges equal to its weight is $O(ve)$.  Therefore, along similar
  lines as the proof of Theorem \ref{unfrm}, the total time complexity
  is $O(v^2e^2)$.
\end{IEEEproof}

Thus, under non-uniform load on all edges, the time complexity of LGLS
is the same as the time complexity of GP (see
\cite{brar_blough_santi__computationally_efficient}, Section 4).

\section{Performance Results}
\label{simul_sec}

\begin{figure}[htbp]
\centering
\includegraphics[width=3.6in]{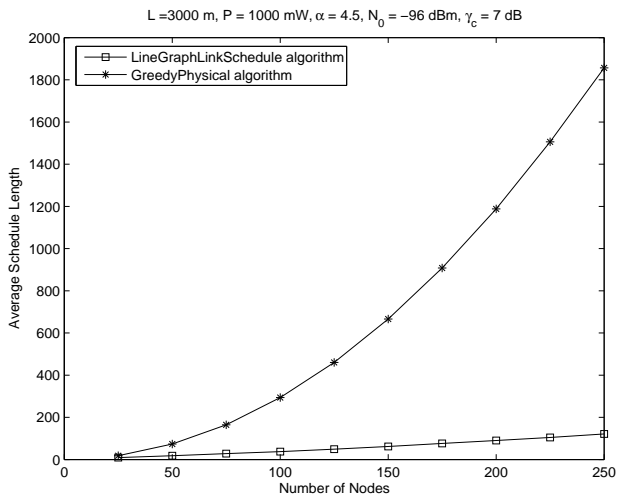}
\caption{Schedule length vs. number of nodes for LGLS and GP algorithms.}
\label{simul}
\end{figure}

In our simulations, $N$ nodes are scattered uniformly in a square by
choosing the $X$ and $Y$ coordinates uniformly from $[0,L]$, where
$L=3000$ m.  The system parameters are $\alpha=4.5$, $\gamma_c=7$ dB,
$P=1000$ mW and $N_0=-96$ dBm, which are chosen to match IEEE 802.11b
operation values \cite{clark_leung__outdoor_ieee}.  The LGLS algorithm
is compared with the GP algorithm
\cite{brar_blough_santi__computationally_efficient} with a weight of
$1$ on all the edges.  We vary $N$ from 25 to 250 in steps of 25.  The
schedule length is averaged over 200 random graphs for each value of
$N$.  Fig. \ref{simul} plots the average schedule length versus the
number of nodes.  We observe that LGLS achieves 50-93\%
lower schedule length than GP.

\section{Conclusions}
\label{conc}
In this paper, we have proposed a novel scheduling algorithm based on
a line graph representation of STDMA network under the physical
interference model.  Our results demonstrate that the schedule length
for the proposed algorithm is substantially lower than that of the
GreedyPhysical algorithm.  This is due to the fact that we have
embedded SINR feasibility conditions into the edge weights of the line
graph, and consequently determined a conflict-free schedule.

\bibliographystyle{IEEEtran}


\bibliography{IEEEabrv,l}

\begin{thebibliography}{10}
\providecommand{\url}[1]{#1}
\csname url@samestyle\endcsname
\providecommand{\newblock}{\relax}
\providecommand{\bibinfo}[2]{#2}
\providecommand{\BIBentrySTDinterwordspacing}{\spaceskip=0pt\relax}
\providecommand{\BIBentryALTinterwordstretchfactor}{4}
\providecommand{\BIBentryALTinterwordspacing}{\spaceskip=\fontdimen2\font plus
\BIBentryALTinterwordstretchfactor\fontdimen3\font minus
  \fontdimen4\font\relax}
\providecommand{\BIBforeignlanguage}[2]{{%
\expandafter\ifx\csname l@#1\endcsname\relax
\typeout{** WARNING: IEEEtran.bst: No hyphenation pattern has been}%
\typeout{** loaded for the language `#1'. Using the pattern for}%
\typeout{** the default language instead.}%
\else
\language=\csname l@#1\endcsname
\fi
#2}}
\providecommand{\BIBdecl}{\relax}
\BIBdecl

\bibitem{nelson_kleinrock__spatial_tdma}
R.~Nelson and L.~Kleinrock, ``{Spatial TDMA: A Collision-Free Multihop Channel
  Access Protocol},'' \emph{{IEEE} Trans. Commun.}, vol.~33, no.~9, pp.
  934--944, Sep. 1985.

\bibitem{gupta_kumar__capacity_wireless}
P.~Gupta and P.~R. Kumar, ``{The Capacity of Wireless Networks},'' \emph{{IEEE}
  Trans. Inf. Theory}, vol.~46, pp. 388--404, Mar. 2000.

\bibitem{ramanathan_lloyd__scheduling_algorithms}
S.~Ramanathan and E.~L. Lloyd, ``{Scheduling Algorithms for Multihop Radio
  Networks},'' \emph{{IEEE/ACM} Trans. Netw.}, vol.~1, no.~2, pp. 166--177,
  Apr. 1993.

\bibitem{wang_li_song__efficient_interference}
W.~Wang, Y.~Wang, X.-Y. Li, W.-Z. Song, and O.~Frieder, ``{Efficient
  Interference-Aware TDMA Link Scheduling for Static Wireless Networks},'' in
  \emph{Proc. ACM MobiCom}, Los Angeles, CA, Sep. 2006, pp. 262--273.

\bibitem{alicherry_bhatia__joint_channel}
M.~Alicherry, R.~Bhatia, and L.~Li, ``{Joint Channel Assignment and Routing for
  Throughput Optimization in Multi-radio Wireless Mesh Networks},'' in
  \emph{Proc. ACM MobiCom}, Cologne, Germany, Aug.-Sep. 2005, pp. 58--72.

\bibitem{brar_blough_santi__computationally_efficient}
G.~Brar, D.~M. Blough, and P.~Santi, ``{Computationally Efficient Scheduling
  with the Physical Interference Model for Throughput Improvement in Wireless
  Mesh Networks},'' in \emph{Proc. ACM MobiCom}, Los Angeles, CA, Sep. 2006,
  pp. 2--13.

\bibitem{moscibroda_wattenhofer__complexity_connectivity}
T.~Moscibroda and R.~Wattenhofer, ``{The Complexity of Connectivity in Wireless
  Networks},'' in \emph{Proc. IEEE Infocom}, Barcelona, Spain, Apr. 2006.

\bibitem{gore_karandikar__link_scheduling}
A.~D. Gore, S.~Jagabathula, and A.~Karandikar, ``{On High Spatial Reuse Link
  Scheduling in STDMA Wireless Ad Hoc Networks},'' accepted for publication in
  IEEE Globecom 2007.

\bibitem{gore_karandikar__high_spatial}
A.~D. Gore and A.~Karandikar, ``{On High Spatial Reuse Broadcast Scheduling in
  STDMA Wireless Ad Hoc Networks},'' in \emph{Proc. National Conference on
  Communications}, Kanpur, India, Jan. 2007, pp. 74--78.

\bibitem{west__graph_theory}
D.~B. West, \emph{Introduction to Graph Theory}, 2nd~ed.\hskip 1em plus 0.5em
  minus 0.4em\relax Prentice Hall, 2000.

\bibitem{jain_padhye__impact_interference}
K.~Jain, J.~Padhye, V.~N. Padmanabhan, and L.~Qiu, ``{Impact of Interference on
  Multi-Hop Wireless Network Performance},'' \emph{{Wireless Networks}},
  vol.~11, no.~4, pp. 471--487, Jul. 2005.

\bibitem{clark_leung__outdoor_ieee}
M.~V. Clark, K.~K. Leung, B.~McNair, and Z.~Kostic, ``{Outdoor IEEE 802.11
  Cellular Networks: Radio Link Performance},'' in \emph{Proc. IEEE ICC},
  Apr.-May 2002, pp. 512--516.

\end{thebibliography}

%

\end{document}